\documentclass{sig-alternate-2013}

\usepackage{tabularx}
\usepackage{texdraw}
\usepackage{graphics}
\usepackage{times}
\usepackage{balance}
\usepackage{url}
\usepackage{subfigure}
\usepackage{graphicx}
\usepackage{epstopdf}
\usepackage{multicol}
\usepackage{amssymb}% http://ctan.org/pkg/amssymb
\usepackage{pifont}% http://ctan.org/pkg/pifont

\usepackage{enumitem}
\usepackage[toc, page] {appendix}

\usepackage{color}
\newcounter{NumberOfComments}
\stepcounter{NumberOfComments}
\usepackage[usenames,dvipsnames]{xcolor}

\permission{Copyright is held by the International World Wide Web Conference Committee (IW3C2). IW3C2 reserves the right to provide a hyperlink to the author's site if the Material is used in electronic media.}
\conferenceinfo{WWW'14,}{April 7--11, 2014, Seoul, Korea.} 
\copyrightetc{ACM \the\acmcopyr}
\crdata{978-1-4503-2744-2/14/04. \\
Include the http://DOI string/url which is specific for your submission and included in the ACM rightsreview confirmation email upon completing your IW3C2 form}

\makeatletter
\def\@copyrightspace{\relax}
\makeatother

\begin{document}

\title{Recommending Investors for Crowdfunding Projects \\  \Large $[$Please cite the WWW'14 version of this paper$]$}
\title{Recommending Investors for Crowdfunding Projects}


\numberofauthors{3} %  in this sample file, there are a *total*
\author{
% 1st. author
%Version as of \/ \today
%Paper ID: 397
\alignauthor
Jisun An\thanks{This work was conducted when the author was in University of Cambridge.}\\
       \affaddr{Qatar Computing Research Institute}\\
       \email{jan@qf.org.qa}
% 2nd. author
\alignauthor
Daniele Quercia\\
       \affaddr{Yahoo Labs, Barcelona}\\
       \email{dquercia@acm.org}
% 3rd. author
\alignauthor 
Jon Crowcroft\\
       \affaddr{University of Cambridge}\\
       \email{Jon.Crowcroft@cl.cam.ac.uk}
}

% There's nothing stopping you putting the seventh, eighth, etc.
% author on the opening page (as the 'third row') but we ask,
% for aesthetic reasons that you place these 'additional authors'
% in the \additional authors block, viz.
%\additionalauthors{Additional authors: John Smith (The Th{\o}rv{\"a}ld Group,
%email: {\texttt{jsmith@affiliation.org}}) and Julius P.~Kumquat
%(The Kumquat Consortium, email: {\texttt{jpkumquat@consortium.net}}).}
% Just remember to make sure that the TOTAL number of authors
% is the number that will appear on the first page PLUS the
% number that will appear in the \additionalauthors section.

\maketitle

\begin{abstract}
\vspace*{-1mm}
To bring their innovative ideas to market, those embarking in new ventures have to raise money, and, to do so, they have often resorted to banks and venture capitalists. Nowadays, they have an additional option: that of crowdfunding. The name refers to the idea that funds come from a network of people on the Internet who are passionate about supporting others' projects. One of the most popular crowdfunding  sites is Kickstarter. In it, creators post descriptions of their projects and advertise them on social media sites (mainly Twitter), while investors look for projects to support. The most common reason for project failure is the inability of founders to connect with a sufficient number of investors, and that is mainly because  hitherto there has not been any automatic way of matching creators and investors. We thus set out to propose different ways of recommending investors found on Twitter for specific Kickstarter projects. We do so by conducting hypothesis-driven analyses of pledging  behavior and translate the corresponding findings into different recommendation strategies. The best strategy achieves, on average, 84\% of accuracy in predicting a list of potential investors' Twitter accounts for any given project. Our findings also produced key insights about the whys and wherefores of investors deciding to support innovative efforts. 
\end{abstract}

\vspace*{-2mm}
%A category including the fourth, optional field follows...
%\category{D.2.8}{Software Engineering}{Metrics}[complexity measures, performance measures]
\vspace*{-2mm}

\keywords{Kickstarter, Twitter, Crowdfunding, Recommending systems} % NOT required for Proceedings

\mbox{ }
\section{Introduction}

\noindent Kickstarter is a crowdfunding website where a founder proposes a project (e.g., smart watch, documentary, video game) and asks the Internet crowd for money. Its use has been growing exponentially: ``The amount pledged on Kickstarter alone grew from \$28m in 2010 to \$320m in 2012''~\cite{economist@2012}.
%~\footnote{http://www.economist.com/blogs/economist-explains/2013/05/economist-explains-unfair-fair-famous-people-kickstarter}. 

However, not all projects are successfully financed. One of the most common reasons for failure is that project founders fail to build a community around them and attract investors. A recent study has indeed found that ``the majority of failed project creators cited the inability to successfully leverage an online audience as a main reason for failing.''~\cite{hui@cscw2014}. That is why we set out to propose automatic ways of matching Kickstarter founders with Twitter investors.  In so doing, we make the following main contributions:

\begin{figure}[t!] \begin{center}
\includegraphics[width=.45\textwidth]{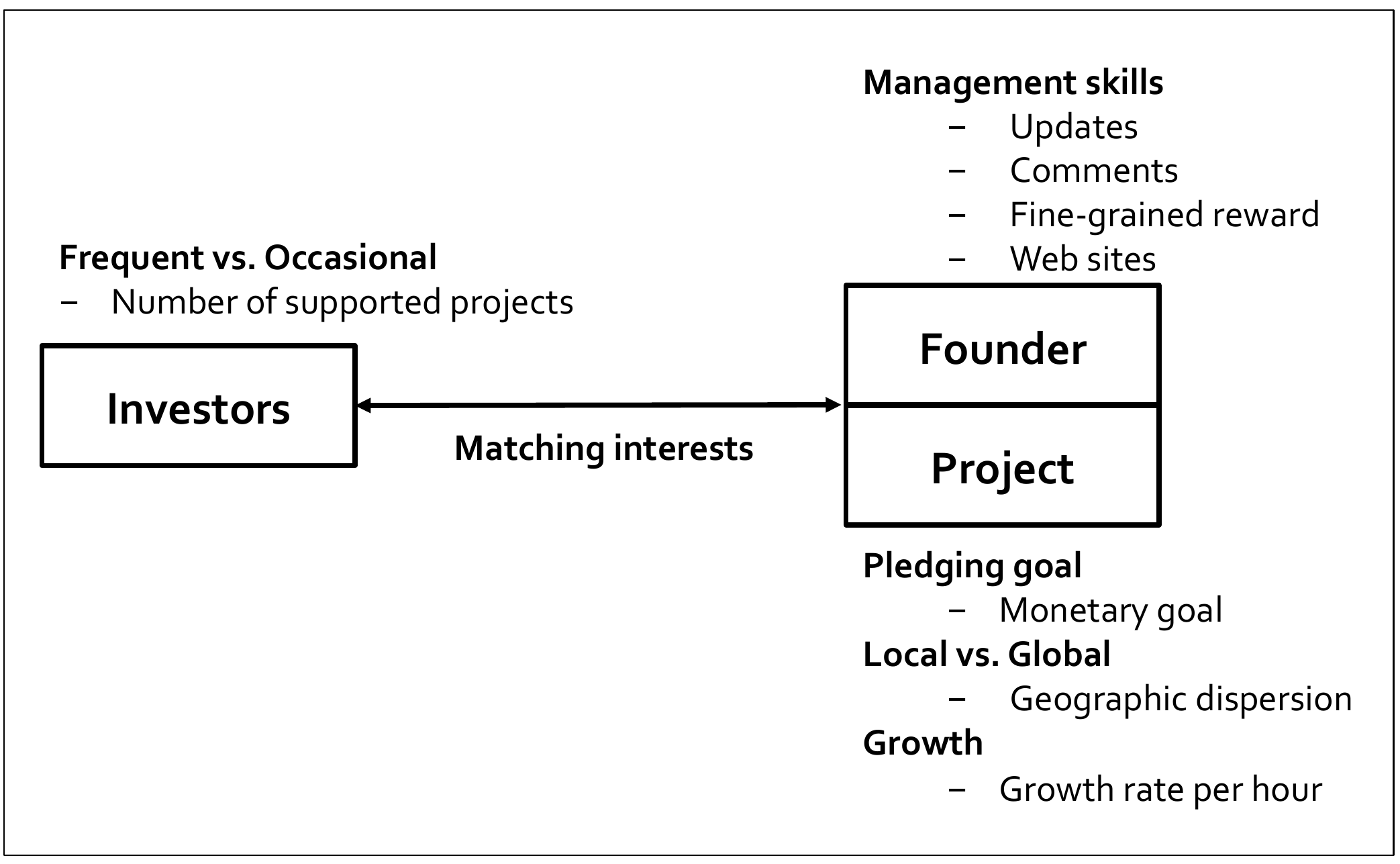}
\end{center} \vspace*{-6mm}
\caption{Aspects hypothesized to affect pledging behavior. We have three actors: (frequent vs. occasional) investors, a founder (with specific project management skills), and the project itself.}
\vspace*{-1mm}\label{fig:framework} \end{figure}

\begin{itemize}
\item We derive a set of well-grounded hypotheses related to pledging behavior (Section \ref{sect:framework}).

\item We crawl data from Kickstarter, including detailed project descriptions and lists of investors, for a period of 3 months  (Section \ref{sect:dataset}). Also, since we need to recommend investors from Twitter, we gather all the tweets related to the projects we previously crawled.   

\item Upon those two datasets, we test a set of of the hypotheses (Section~\ref{sect:analysis}). We find that investors behave differently depending on whether they are frequent or occasional supporters on the site. As opposed to occasional investors (51\% of the investor base supported less than 4 projects), frequent ones (11\% supported more than 32 projects) tend to fund efforts that are well-managed and match their own interests. By contrast, occasional investors pay less  attention to any of those aspects and thus act as donors, mainly on art-related projects. 

\item Upon the quantitative analysis, we build a statistical model to recommend potential investors from Twitter (Section \ref{sect:prediction}). Our model achieves 84\% of accuracy in predicting an unordered list of investors, and an average percentile-ranking of 0.32 (i.e., a 36\% gain over the random baseline) in predicting an ordered list. Also, in situations of investor cold start (no previous information about an investor is available), we are still able to predict who funds what with an accuracy of 69\%  from Twitter-derived activity features, and an average percentile-ranking of 0.40 (20\% gain over the random baseline).

\end{itemize}

We conclude by discussing the theoretical and practical implications of our findings (Section~\ref{sect:discussion}).

\section{Related work}
\label{sect:related_work}

\noindent The first crowdfunding site started in 2001 and now there are more than 450 of such sites. They together have raised \$2.8 billion and successfully funded more than 1 million projects only in 2012~\cite{massolution@2013}. Kickstarter is the largest site in USA. In it, a founder proposes a project by posting information about the project's  purposes, monetary goals,  time left to reach those goals, the way funds will be used, and potential rewards (e.g., the founder might offer a signed CD in exchange of a donation). To increase his/her likelihood of success, the founder usually posts videos and pictures to visually explain the project. (S)he also connects the project's page to dedicated social-networking accounts~\cite{mollick@ssrn2012}. As crowdfunding sites have emerged, small entrepreneurs without an access to traditional venture capital's fundings have benefited from this new source of cash flow.

Crowdfunding has recently attracted the attention of researchers in various disciplines, from business and economics to computer science.  Economists have investigated pleading behavior and they, for example,  found that crowdfunding eliminates distance-related economic frictions, yet initial findings tend often to come from family, friends and acquaintances~\cite{agrawal@2011}.

Most of the work by computer scientists has focused, instead, on predicting whether projects will be successfully funded or not.  Mollick found that variables under two categories - preparedness (e.g., existence of video, spelling check, number of updates) and social capital (e.g., number of the founder's Facebook friends) - are strongly related to the success of a project~\cite{mollick@ssrn2012}. Greenberg \textit{et al.} found that SVM could predict, at the time of launch, whether a project will fail or succeed with a roughly 68\% accuracy~\cite{greenberg@chi2013}. More recently, based only on the use of language in project descriptions, Gilbert \emph{et al.} were able to predict failure or success - they indeed found that there are specific phrases that are powerful predictors of success~\cite{gilbert@cscw2014}. These phrases are mainly related to six general persuasion principles: 1) reciprocity, 2) scarcity, 3) social proof, 4) social identity, 5) liking, and 6) authority. After launch, one could also track features that change as the project evolves. In this vein, upon the time series of money pledged and tweets, Etter \textit{et al.} were able to predict success/failure with an accuracy of 85\% at early stages - that is, just after 15\% of the entire duration of a campaign has passed~\cite{etter@cosn2013}.
 
Hui \textit{et al.}  conducted a throughout qualitative analysis based on 45 interviews and found that the work behind setting up a crowdfunding project unfolds in five main steps:  1) prepare; 2) test; 3) publicize; 4) follow through; and 5) contribute. They then went on recommending which tools computer scientists could build  for supporting each of the steps~\cite{hui@cscw2014}. The most difficult step was identified to be the third one:  founders repeatedly failed to build a community and attract potential investors: ``The majority of failed project creators cited the inability to successfully leverage an online audience as a main reason for failing.''~\cite{hui@cscw2014} \\

\noindent Based on this literature review, we might conclude that an automatic way of matching projects with investors is needed. To propose such a way, we carry out an analysis that unfolds in three steps:  derive few hypotheses concerning pledging behavior; 2) collect data from Kickstarter and Twitter to test those hypotheses; and 3) based on the findings, propose and evaluate models that match projects with potential investors.

\begin{table*}[t!] \begin{center} \small\frenchspacing
\begin{tabular}{l} 
\hline
\textbf{Hypothesis}  \\
\hline
 \emph{[H1] A project is likely to be financed by frequent investors if its  founder:}  \\ 
 ~~~~~~~~\emph{[H1.1] frequently updates the project after launching it.}   \\ 
 ~~~~~~~~\emph{[H1.2] answers the potential investors' requests.}   \\ 
 ~~~~~~~~\emph{[H1.3] allows for fine-grained funding levels.} \\ 
 ~~~~~~~~\emph{[H1.4] sets a dedicated web site.}  \\ 
 \emph{[H2]  A project with a high goal is likely to be financed by frequent investors.}   \\
\emph{[H3] A local project is likely to be supported by occasional investors.}     \\
\emph{[H4] A fast-growing project is likely to be financed by frequent investors.}   \\
\emph{[H5] Active investors tend to fund projects that match their own interests.}    \\
\hline
\end{tabular} \end{center} \caption{List of Hypotheses in this Study. }
\label{tab:summary_hypothesis} \end{table*}

\section{Investors vs. Donors}
\label{sect:framework}

\noindent Pledging behavior might well differ from one investor to another. It has been found that 20-40\% of \emph{initial} fundings in Kickstarter come from family and friends~\cite{economist@2012-2}. These individuals tend to be newcomers or occasional investors who support projects because of their personal relationships with the founders. By contrast, users who are very active in Kickstarter  are passionate about the community and fund a project for different reasons. To test the extent to which this distinction impacts pledging behavior, we differentiate investors depending on whether they are occasional (they have supported, say, less than 4 projects) or frequent (they have supported more than 32 projects), and formulate a set of hypotheses, which Table~\ref{tab:summary_hypothesis} collates for convenience. We expect that the more active a supporter has been, the more (s)he will behave as an investor, and the less as a friend donating money. More specifically, as opposed to occasional investors, frequent ones are expected to: \\

\noindent \textbf{Pay Attention to Founder Skills}. Successful venture founders are good managers as well: ``Many entrepreneurs make the mistake of thinking that venture capitalists are looking for good ideas when, in fact, they are looking for good mangers in particular industry segments.''~\cite{zider@hbr1998} We expect that frequent investors will pay attention to the way a project is managed. Since management of a Kickstarter project translates into frequent updates after launch and audience interactions, our first hypothesis is: \textit{[H1] A project is likely to be financed by frequent investors if its  founder:} \textit{[H1.1] frequently \textit{updates} the project after launching it} (i.e., (s)he spends extra effort to make it happen~\cite{mullins@hbr2007});  \textit{[H1.2] answers the potential investors' requests} (i.e., she interacts with the audience); \textit{[H1.3] allows for fine-grained funding levels}; and \textit{[H1.4] sets a dedicated web site}. \\

\noindent  \textbf{Invest in ``High Capital'' Projects}. Since Kickstarter follows ``all-or-nothing'' model (i.e., projects that do not reach their pledging goals do not receive a penny), founders tend to set realistic goals for the amount to be raised. It is reasonable to assume that projects with high goals are ambitious (e.g., bringing a new video game to market) and thus tend to be preferentially financed by frequent and experienced investors~\cite{sahlman@hbr1997}: \textit{[H2] A project with a high goal is likely to be financed by frequent investors.} \\

\noindent  \textbf{Invest in Geographically Global Projects}. Since friends and acquaintances tend to be geographically close, we expect that those who support (geographically) local projects are occasional investors, while frequent ones would also support global projects~\cite{agrawal@2011}. We define the geographic dispersion of project $p$ as:
\begin{equation}
{G_p}= \frac{1}{N_p} \sum_{b \in B_p} {{D(l_f,l_b)}}
\end{equation}
which measures the mean distance for all pledges of a project (distance between a project $p$ and an investor $b$). In Kickstarter, founders and investors tend to reveal their location in terms of city. We thus convert city names into  geographic coordinates of the corresponding centroids (latitude and longitude) and measure the Harvesine distance $D$ between the founder's location ($l_f$) and each investor's location ($l_b$). $N_p$ is number of all investors for project $p$. With this definition, low geographic dispersion is associated with projects with investors who live close to the founder, while high dispersion is associated with investors who live far away. The corresponding hypothesis then reads: \textit{[H3] A local project is likely to be supported by occasional investors.} \\

\noindent \textbf{Pay Attention to Fast Growing Projects}. As opposed to occasional investors, frequent ones are familiar with the site and are thus expected to be able to quickly spot fast-growing opportunities~\cite{sahlman@hbr1997,stancill@hbr1981}: \textit{[H4] A fast-growing project is likely to be financed by frequent investors}. \\

\begin{figure}[t!] \begin{center}
{\includegraphics[width=.49\textwidth]{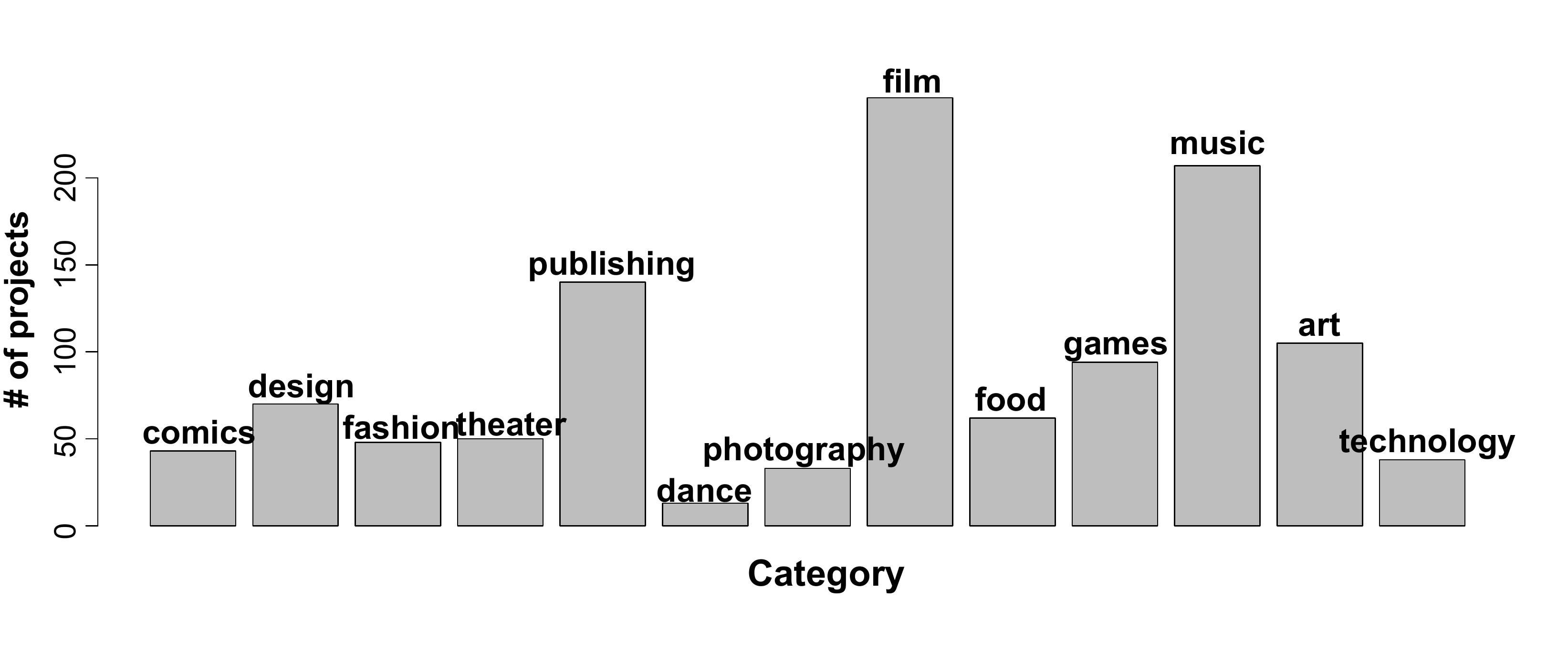}}
\vspace*{-4mm} \end{center} \vspace*{-4mm}
\caption{Frequency distribution for each category.} 
\vspace*{-4mm}\label{fig:hist-category} \end{figure}

\noindent \textbf{Invest in Project Categories of Interest}. When deciding whether to fund a project or not, frequent investors might well be looking for projects that match their own  interests, while occasional investors do not concern that as they, for example, tend to support a friend.  \textit{[H5] Active investors tend to fund projects that match their own interests.}  To capture each investor's interests, we keep track of: 1) which project categories (s)he supports, and 2) the topics classified using LDA topic modeling~\cite{blei@2003,griffiths@2004, steyvers@2007} (s)he mentions on Twitter. \\

\section{Dataset}
\label{sect:dataset}

\noindent Since Kickstarter is not accessible from an API, we need to build a crawler running on University servers on which we run our data analysis as well. We gather all the projects featured on the  \textit{Recently Launched}\footnote{\url{http://www.kickstarter.com/discover/recently-launched}} Kickstarter page between July 2013 and October 2013. Once a new campaign is identified, we crawl its category (e.g., Film, Dance, Art, Design and Technology), funding goal, and deadline. We then regularly check each project's page for any change in the amount of pledged money and total number of investors.  To have a comparable dataset of projects, we have eliminated  345 Kickstarter projects that happened to be outside USA. In so doing, we have collected information about 1,149 projects that were funded by 78,460 investors with a total number of 177,882 pledges. These projects are classified in 13 categories, and the most popular ones are Film, Music, Publishing, and Art (Figure~\ref{fig:hist-category}).

\begin{table}[t!]
\begin{center}
%\small \frenchspacing
\begin{tabular}{llll}
%\hline
 & \textbf{Successful} & \textbf{Failed} & \textbf{Total} \\
\hline
Projects & 520 &  629 & 1,149 \\
Proportion & 45.3\% & 54.7\% & 100\% \\
Investors & - & - & 78,460  \\
Pledges & 148,257 & 29,625 & 177,882\\
Pledged (\$) & 10,517,919 & 1,872,741 & 12,390,660 \\
Tweets & 49,943&  21,372 & 71,315\\
\hline
\end{tabular}
\end{center}
\vspace*{-2mm}
\caption{Statistics for the Kickstarter dataset.}
\label{tab:global_statistics}
\end{table}

\begin{table}[t!]
\begin{center}
%\small \frenchspacing
\begin{tabular}{llll}
%\hline
 & \textbf{Successful} & \textbf{Failed} & \textbf{Total} \\
\hline
Goal (\$) & 11,033.90 & 30,716.86 & 20,875.38 \\
Duration (days) & 28.56 & 29.25 &28.91 \\
Number of investors & 285.11 & 47.09 & 166.10 \\
Pledge (\$) & 79.71 & 60.13 & 68.99\\
Final amount & 168.93\% & 19.51\% & 94.22\% \\
Number of tweets & 101.93 & 44.43 & 73.18 \\
\hline
\end{tabular}
\end{center}
\vspace*{-2mm}
\caption{Statistics for the Kickstarter projects. All reported are the average of each measure for Kickstarter projects.}
\label{tab:project_statistics}
\end{table}

During the same period of time, we collected all tweets\footnote{A server located at Computer Laboratory, University of Cambridge was used to collect Twitter data.} containing the keyword ``\textit{kickstarter}'' from the publicly available Twitter search API. If a tweet matches one of our Kickstarter projects (if a tweet contains project title or shortened url directing to the project page), we match the tweet's content with the project, resulting in a total of 71,315 tweets.

Table~\ref{tab:global_statistics} reports general statistics of our Kickstarter dataset, and Table~\ref{tab:project_statistics} reports statistics specifically about the projects. The numbers in Table~\ref{tab:project_statistics} are the average of all projects. Out of our 1,149 projects, \$12.3M were pledged and 520 projects (45.2\%) were successfully funded (i.e., they met their pledging goals). This success rate is similar to the general one published by Kickstarter itself\footnote{\url{http://www.kickstarter.com/help/stats.} All statistics reported are retrieved on 3rd November 2013.}: 43.85\%.  On average, as opposed to unsuccessful projects, the ones  that are successfully funded tend to have lower financial goals (\$11,033 \emph{vs.} \$30,716); have more investors (285 \emph{vs.} 47), raise more funds than their goals would require (169\% \emph{vs.} 19\%), and generate more tweets (101 \emph{vs.} 44)  (Table~\ref{tab:project_statistics}). In our dataset, 85\% of donations get into successful projects (this is 86\% in Kickstarter). The average duration of a successful campaign is 28.9 days; however, it takes just few days (13 days) to be fully financed. On the other hand, unsuccessful campaigns, which are 54.8\%, take just 19.5\% of the required investment (this was 20\% by Kickstarter in 2012~\cite{economist@2012-3}). Since the previous statistics in our dataset match those in the larger sample, we conclude that our dataset is fairly representative.

\begin{figure}[t!] \begin{center}
\vspace*{-2mm}{\includegraphics[width=.35\textwidth]{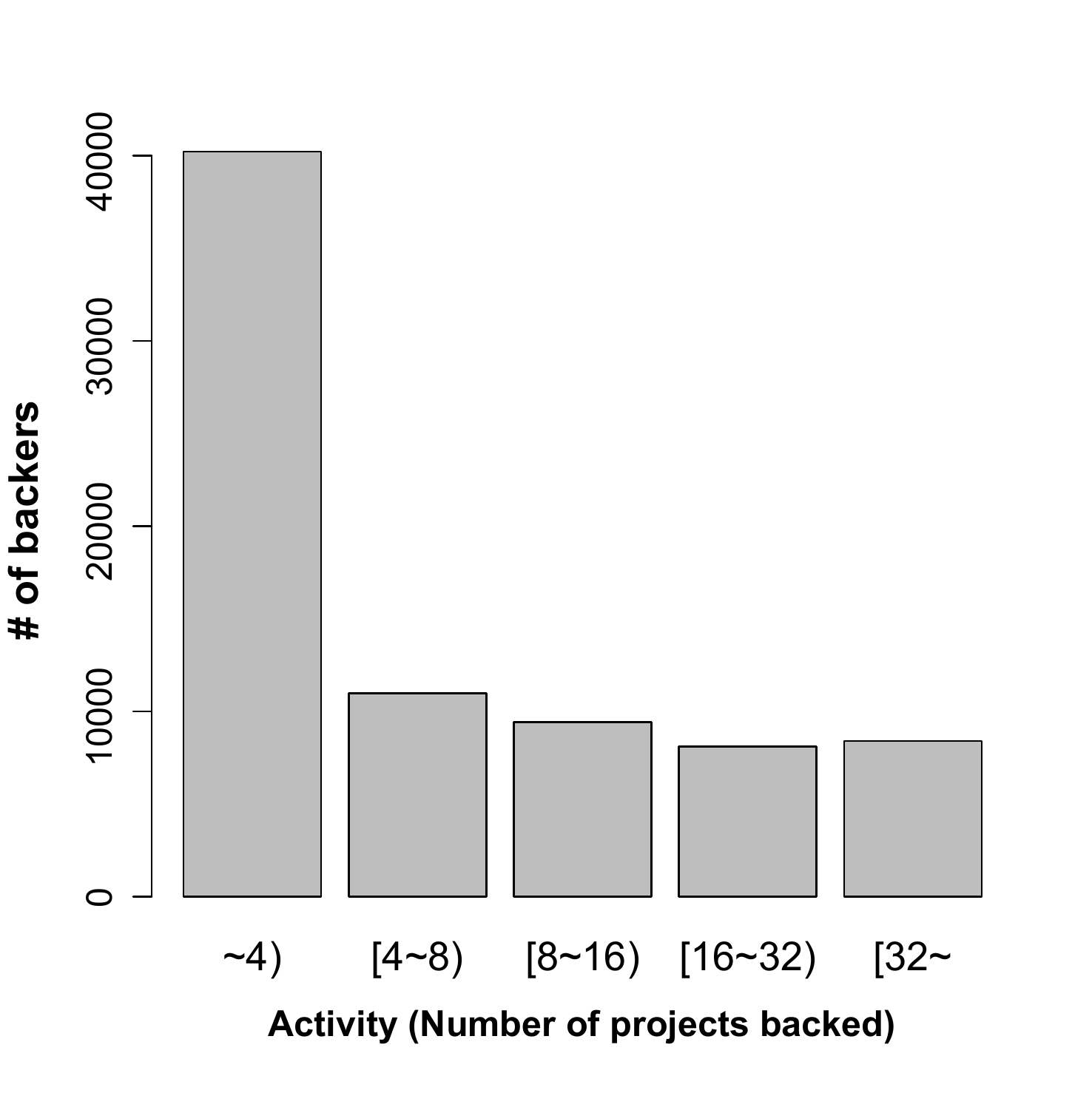}}
\vspace*{-4mm} \end{center} \vspace*{-4mm}
\caption{Distribution of investor activity levels. These levels are quantified using the number of projects supported by each investor.} 
\vspace*{-4mm}\label{fig:hist_activity} \end{figure}

\begin{table}[t!] \begin{center}
{\includegraphics[width=.45\textwidth]{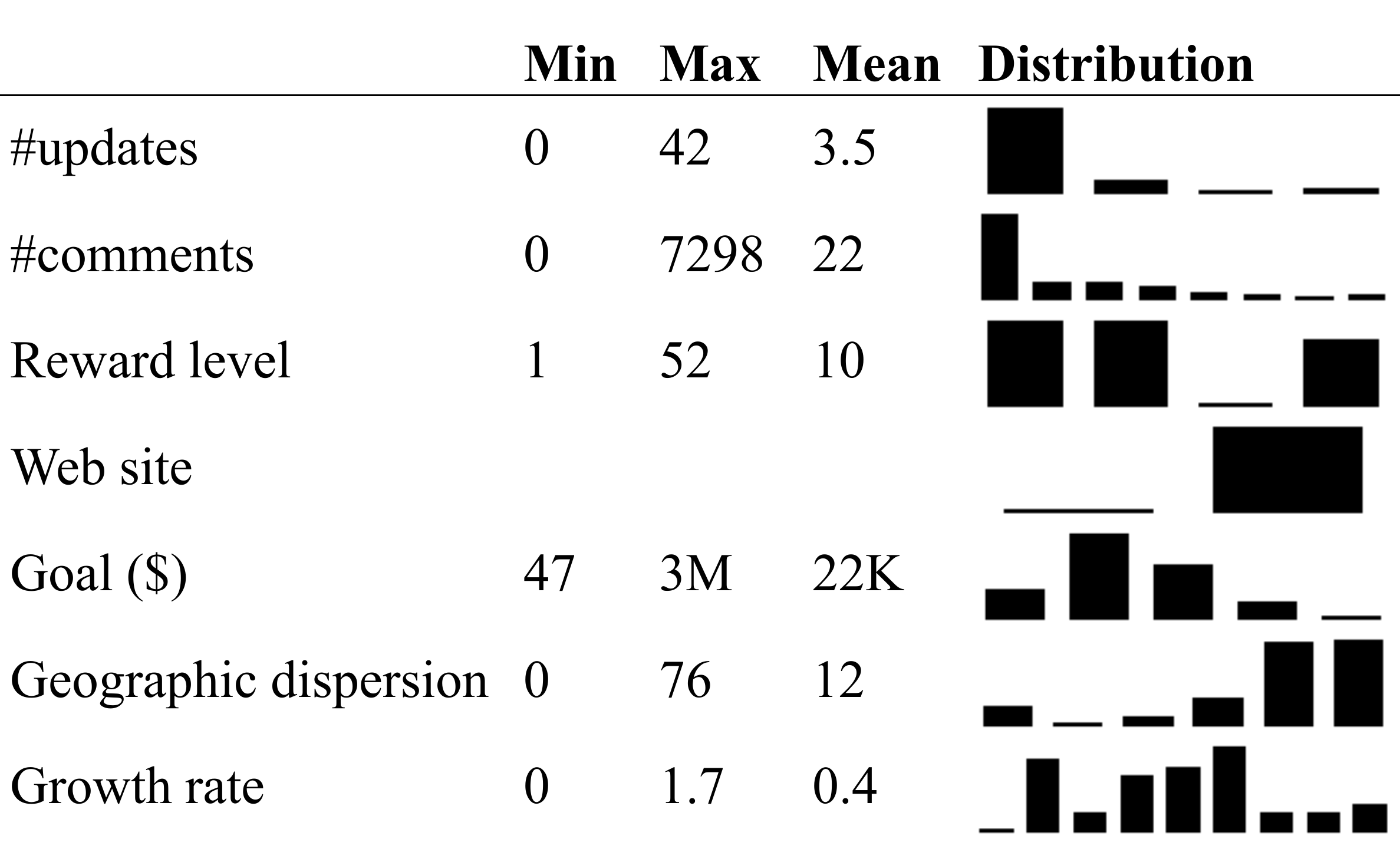}}
\vspace*{-2mm} \end{center} \vspace*{-2mm}
\caption{Our predictive features: minimum, maximum, mean, and frequency distributions (a-h). The $x$-axis reports the values of each feature (which is log-transformed if skewed), and the $y$-axis reports the number of users.}
\vspace*{-3mm}\label{fig:distribution} \end{table}

People who back a project for the first time often go on to back other projects. Among 78,460 people who have backed one of projects in our dataset, 22K (28\%) people have backed two or more projects (Kickstarter has reported 29\% of all backers as repeat backers). On average, investors in our dataset supported three projects. We segment them into two groups - occasional investors who funded less than 4 projects (51\%), and frequent ones who funded more than 32 projects (11\%).  Figure~\ref{fig:hist_activity} displays the frequency distribution of their activity levels. We also display distribution of each project feature in Table~\ref{fig:distribution}. Since the distributions of the features are skewed, we shown their log-transformed distributions if it is necessary.

\begin{figure*}[ht!] \begin{center}
\subfigure[{[H1.1]} \#updates]{\includegraphics[width=.32\textwidth]{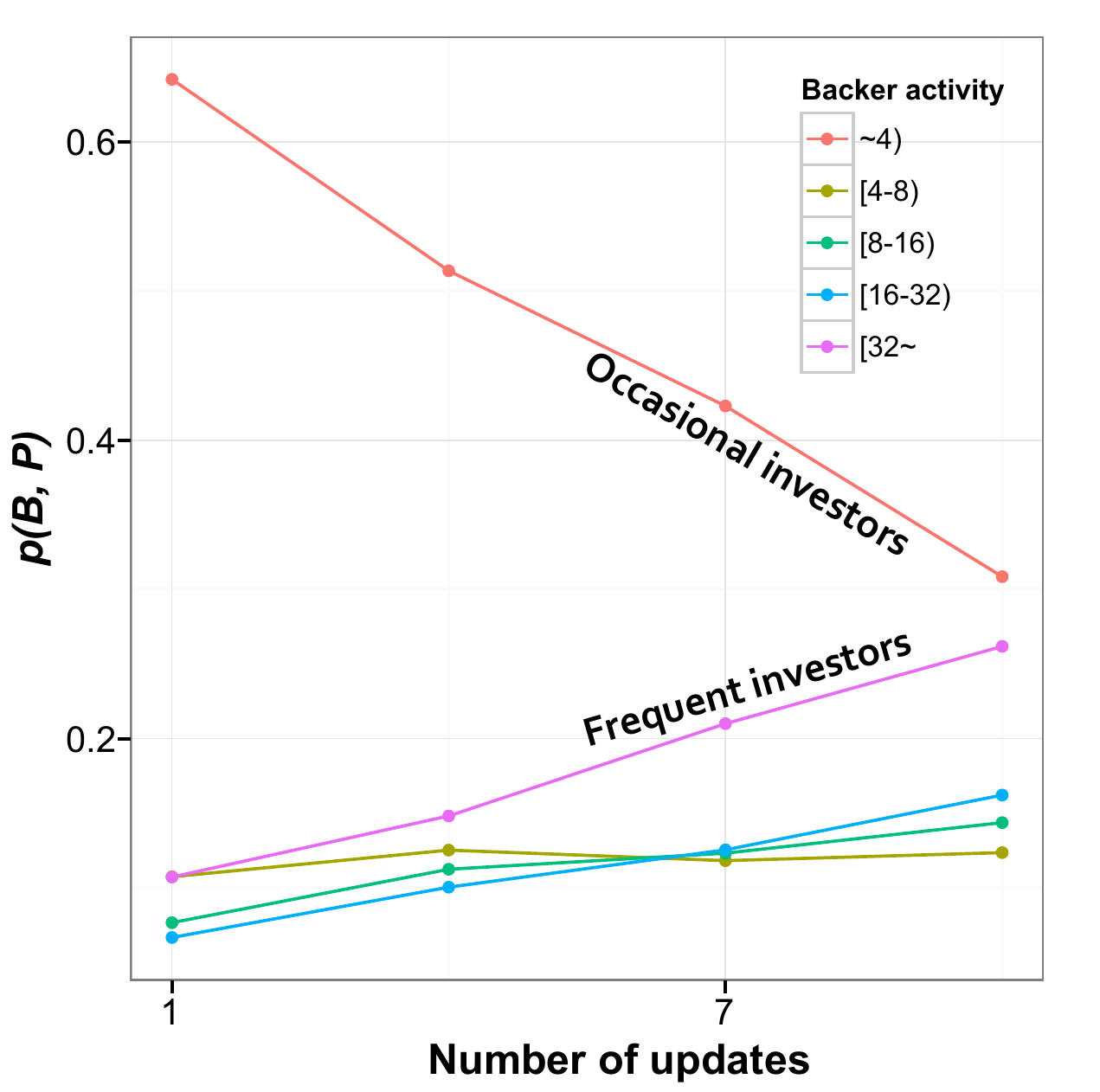} \label{fig:group_numbacked_all_prob_numupdates}}
\subfigure[{[H1.2]} \#comments]{\includegraphics[width=.32\textwidth]{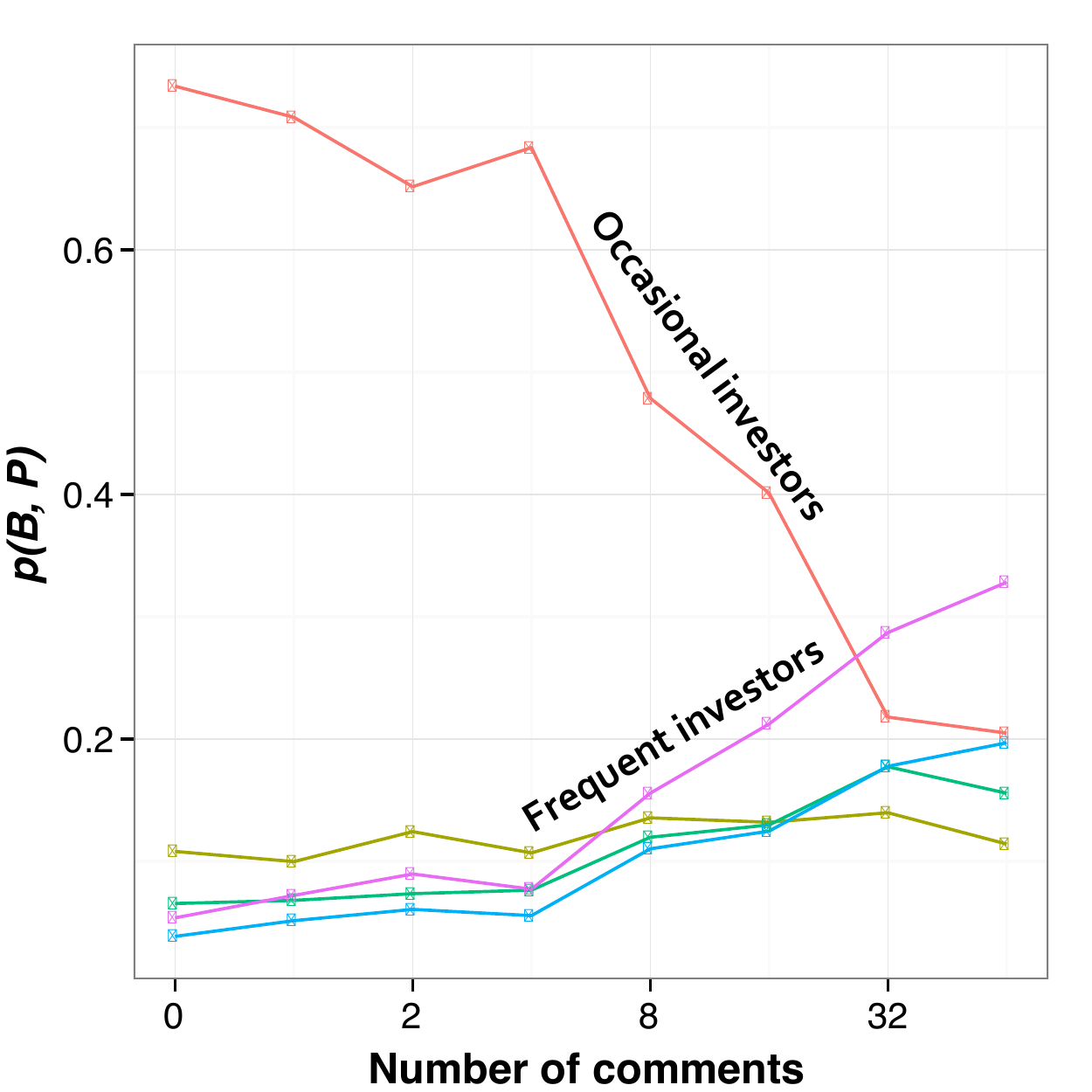} \label{fig:group_numbacked_all_prob_numcomments}}
\subfigure[{[H1.3]} Reward level]{\includegraphics[width=.32\textwidth]{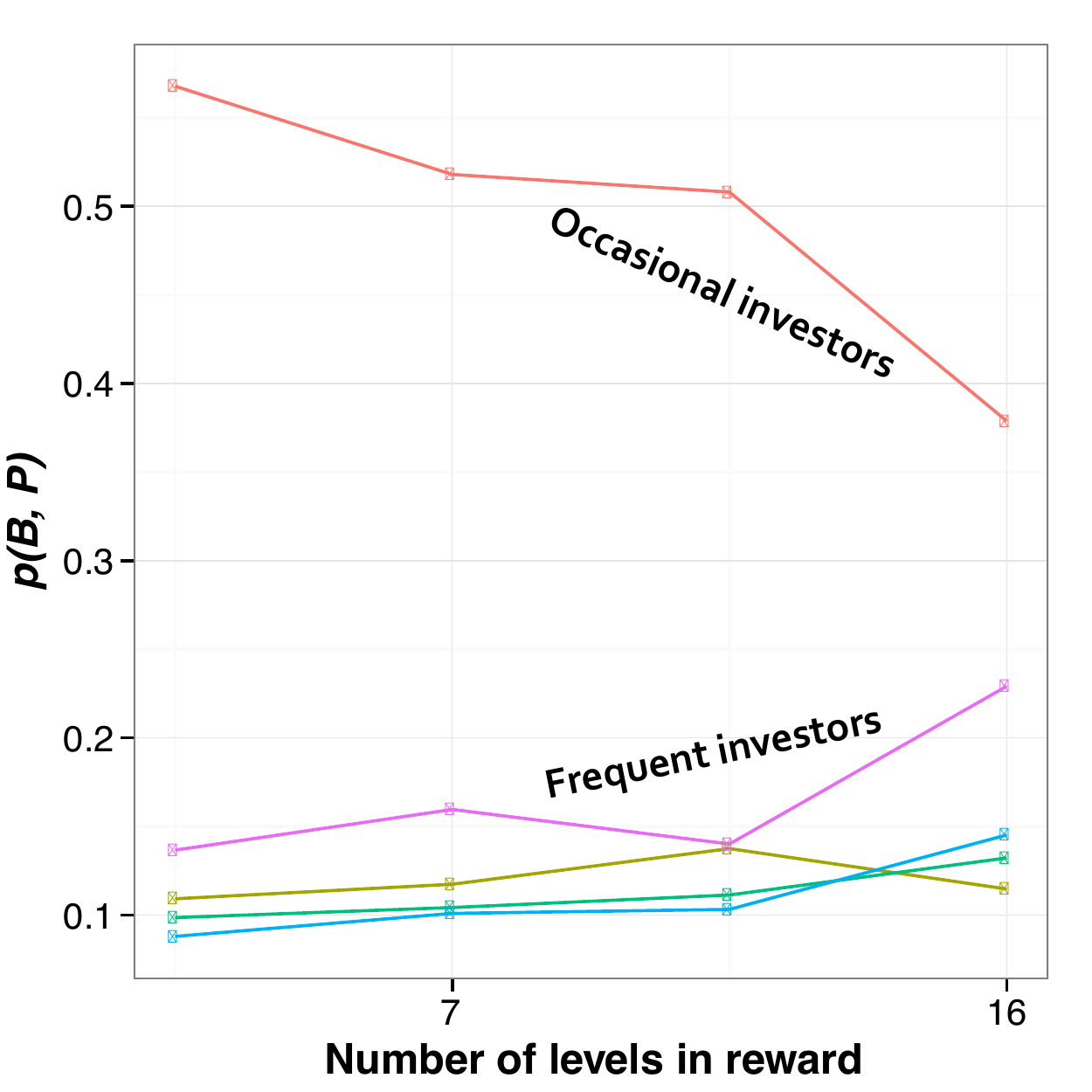} \label{fig:group_numbacked_all_prob_rewardlevel}}
\subfigure[{[H2]} Goal (\$)]{\includegraphics[width=.315\textwidth]{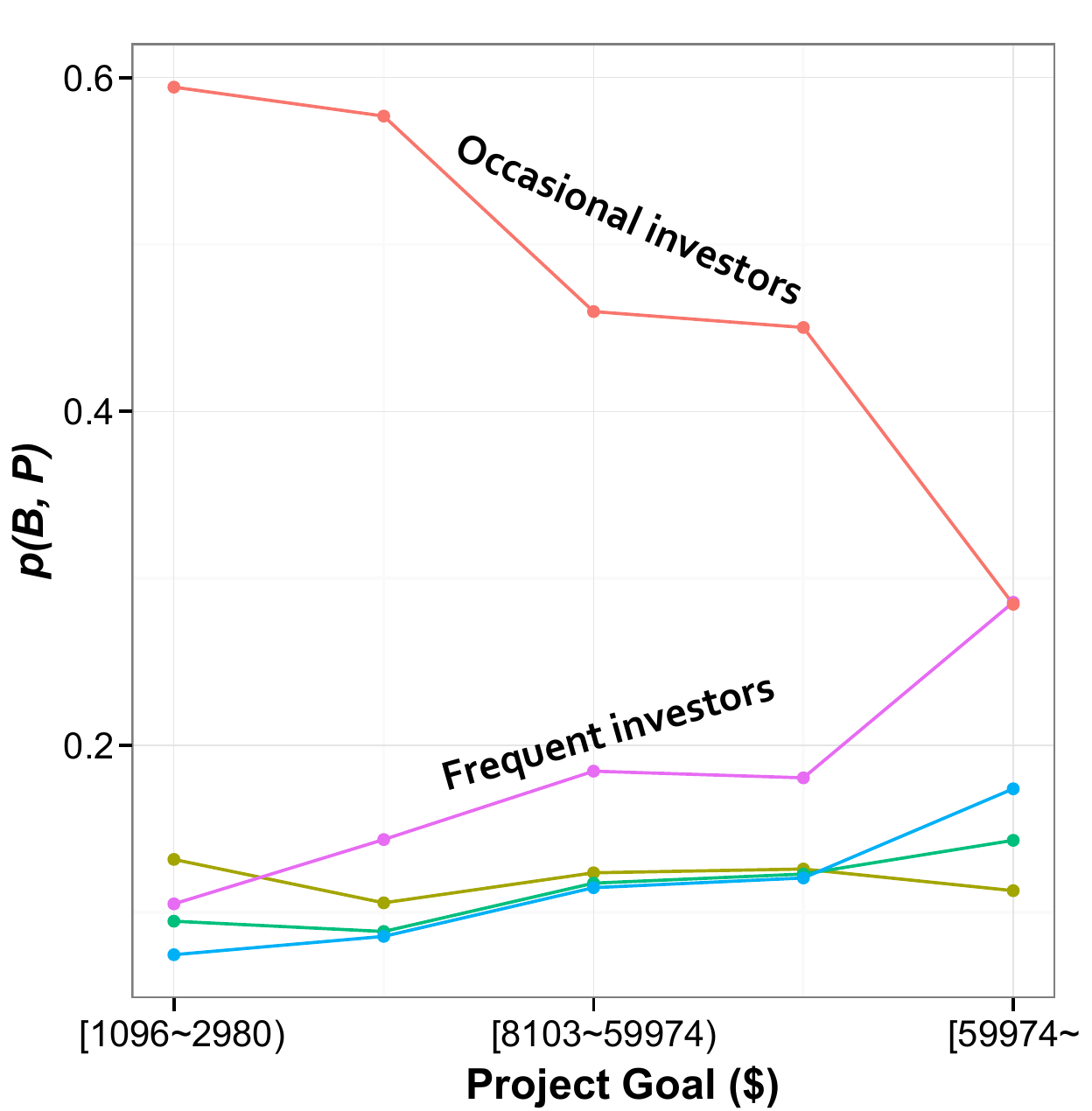} \label{fig:group_numbacked_all_prob_goal}}
\subfigure[{[H3]} Geographic dispersion]{\includegraphics[width=.32\textwidth]{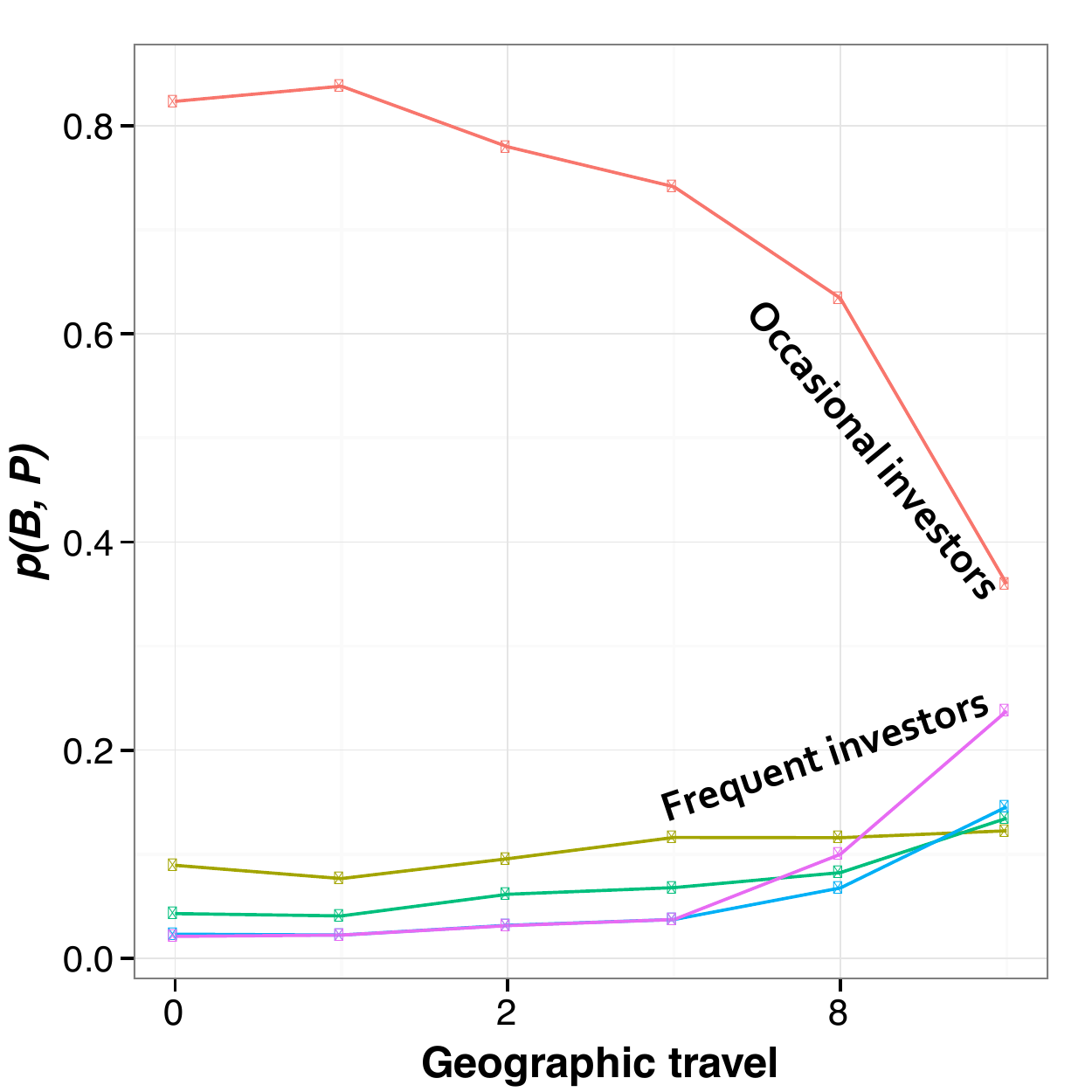} \label{fig:group_numbacked_all_prob_geo_span}}
\subfigure[{[H4]} Growth rate]{\includegraphics[width=.32\textwidth]{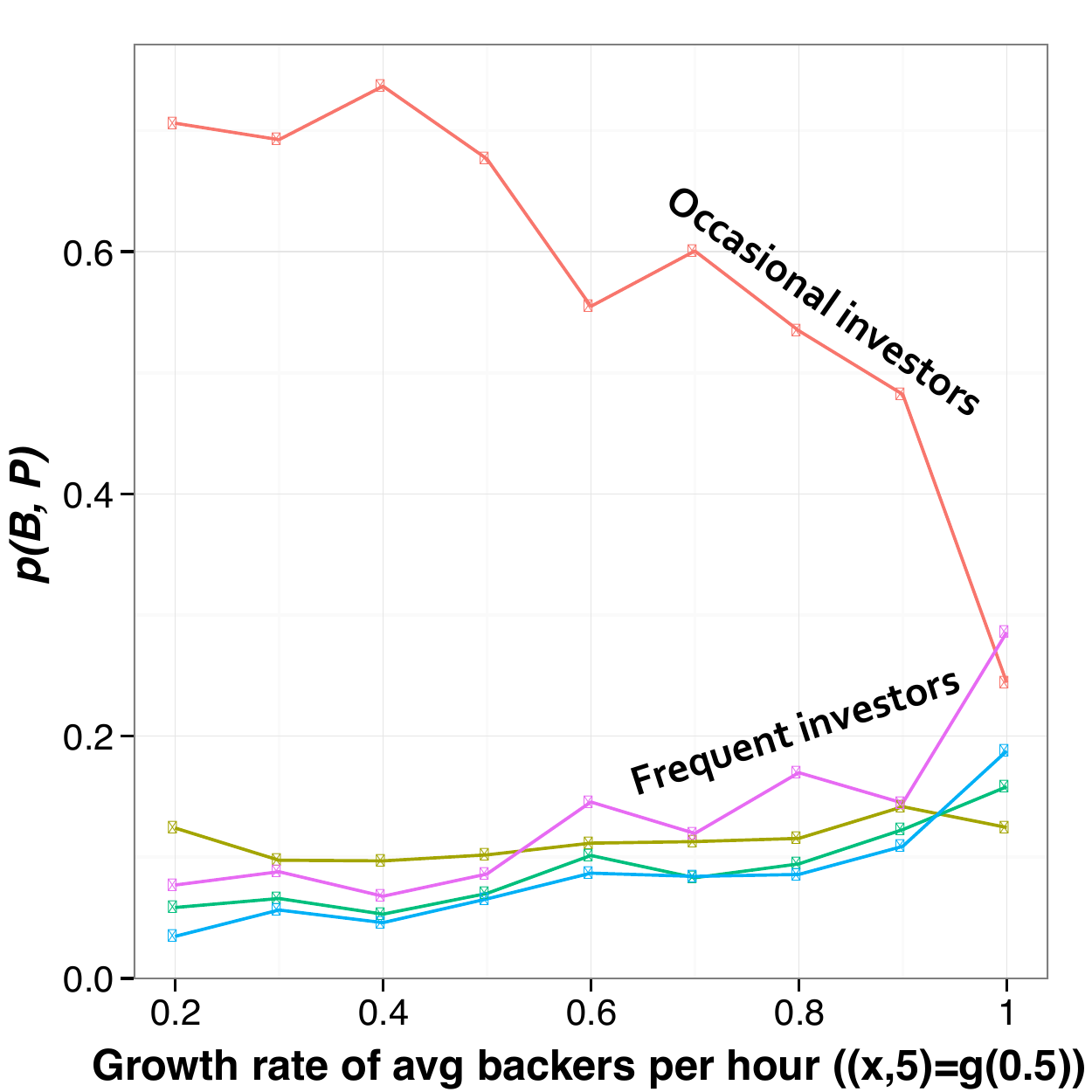} \label{fig:group_numbacked_all_prob_growthrate}}
\end{center} \vspace*{-2mm}
\caption{Probability of investor $B$ to fund project $P$, as a function of \emph{(a)} the number of updates made by the project's founder; \emph{(b)} the number of comments received by the project;  \emph{(c)} the rewards levels made available to investors; \emph{(d)} the project's pledging goal; \emph{(e)} the investors' geographic span; and \emph{(f)} the project's growth rate.} 
\vspace*{-1mm}\label{fig:prob_all_numinvestors} \end{figure*}

\section{Pledging Behavior}
\label{sect:analysis}

\noindent Having this dataset at hand, we are now able to quantitatively analyze investors' pledging behavior. To do so, we will often resort to the probability that a investor of type $B$ (e.g., occasional) will fund a project of type $P$ (e.g., projects of smart watches): 

\begin{equation}
\textrm{p}(B | P ) = {\frac{p(B \cap P )}{p(P)}}
\end{equation}

\noindent We compute this probability by counting the fraction of investors of type $B$ who funded projects of type $P$ (e.g., occasional investors who funded projects of smart watches) out of all  investors who backed projects of type $P$ (e.g., investors of any type who  funded  projects of smart watches). When testing our hypotheses, we will compute the probability of funding a project $P$ for different investor types, and type is defined depending on the level of pledging activity: from occasional investors who supported less than four projects, to investors who supported less than 8, up to frequent investors who  supported more than 32. One of the probabilities $\textrm{p}(B | P )$ could then be $\textrm{p}(occasional | P )$, which is  the probability  of an occasional investor to fund project $P$.  After clarifying our notation, we are now ready to test each of the main five hypotheses one by one.

\mbox{ } \\
\noindent \textit{[H1]}  \textit{A project is likely to be financed by frequent investors if the project  founder:} 
\begin{description}
\item \textit{[H1.1] frequently \textit{updates} the project after launching it};  
\item \textit{[H1.2] answers the potential investors' requests}; 
\item \textit{[H1.3] allows for fine-grained funding levels}; 
\item \textit{[H1.4] sets a dedicated web site}.
\end{description}

We find that frequent investors are more likely to pledge projects with frequent updates (Figure~\ref{fig:group_numbacked_all_prob_numupdates}) and higher level of founder engagement (number of comments) (Figure~\ref{fig:group_numbacked_all_prob_numcomments}). As the number of comments increases by an order of 2, the pledging probability increases by 10\%. Funding levels (Figure~\ref{fig:group_numbacked_all_prob_rewardlevel}) and dedicated web sites do matter, but to a lesser extent compared to the previous two features. We also compute the Pearson's correlation coefficients between a management strategy and investor's activity levels. These coefficients range from -1 (strongest negative correlation) to 1  (strongest positive correlation), and are 0 when there is no correlation. Frequent investors seem to decide whether to fund a project or not depending on the number of updates done by the founder  ($r = 0.26, p < 0.05$) and on the number of  comments the project has received ($r = 0.19, p < 0.05 $). The presence of different reward levels ($r = 0.05, p < 0.05$) and of a dedicated web site ($r = 0.10, p < 0.05$) are considered by occasional and frequent investors alike. Overall, the first hypothesis is supported.

\mbox{ } \\
\noindent
\textit{[H2] A project with high pledging goal is likely to be financed by frequent investors.} 

By dividing projects into  5 categories depending on their pleading  goals, we find that, the higher a project's goal, the less likely occasional investors support it. By contrast, frequent investors are more likely to fund high-goal projects (Figure~\ref{fig:group_numbacked_all_prob_goal}). The correlation between investor activity and the pledging goals of supported projects is indeed positive: $r = 0.21, p < 0.05$. Hence the second hypothesis is also confirmed. 
 From the perspective of a recommender system that matches investors with projects, this result suggests that high-goal projects should be preferentially matched with frequent investors.

\mbox{ } \\
\noindent \textit{[H3] A local project is likely to be financed by occasional investors.} 

As we mentioned in Section~\ref{sect:framework}, we compute the average geographic span of a project's investors to measure the extent to which a project attracts local \emph{vs.} global fundings. We then plot the pledging probability as a function of geographic span (Figure~\ref{fig:group_numbacked_all_prob_geo_span}) and find that occasional investors largely fund  projects with low geographic span (i.e., local projects), while frequent investors fund projects with high span.  The correlation coefficient between investor activity and geographic span is $r = 0.32, p < 0.05 $, and that supports the third hypothesis. For a recommender system, this result means that local projects should be matched with local Kickstarter users who tend to be occasional investors.

\mbox{ } \\
\noindent \textit{[H4] A fast-growing project is likely to be financed by frequent investors}. 
   
We confirm this hypothesis as well since we find that frequent investors tend to indeed support high-growth projects (Figure~\ref{fig:group_numbacked_all_prob_growthrate}). By contrast, occasional investors  do not select the projects to support depending on growth rate - they just happen to support the majority of projects that are characterized by limited growth. The correlation between investor activity and project growth is positive and is $r = 0.17, p < 0.05$.

\mbox{ } \\
\noindent \textit{[H5] Frequent investors tend to support projects that match their own interests.}

To test this hypothesis, we  consider investors who are on Twitter and crawl 200 tweets (at  most) for each of them using the Twitter Public API. To compute the topical similarity between a project's description and an investor's tweets, we run the topic model Latent Dirichlet Allocation (LDA) on the tweets and the project descriptions. As a result, each project is represented by a topic vector, and each investor's Twitter account is represented by another topic vector. To assess whether a project's description matches an investor's interests, we simply compute the cosine similarity between the project's topical vector and the investor's. We do so for all project-investor pairs and find that  frequent investors do indeed support projects that match their own interests, while occasional investors' topical interests are not really matching to projects they supported. We also find that frequent investors fund projects in a variety of categories, while occasional ones tend to stick with the same category, if they happen to fund more than one project. The correlation between investor activity and project-investor cosine similarity is positive: $r = 0.20, p < 0.05$. In practice, this suggests that topical matching between projects and investors tends to work better for frequent investors than for occasional ones.

\subsection{Summary}
\noindent We find that frequent investors are likely to fund projects that are well-managed; have high pledging goals; are global; grow quickly; and match their interests. Occasional investors, instead, do not seem to base their decisions on those aspects. We might thus infer that those who have supported a considerable number of projects  act in ways similar to how investors would do, while occasional supporters appear to be behaving as charitable donors. As hinted by a news article~\cite{economist@2012-2}, we suspect that  occasional investors are lured into Kickstarter by their own friends and family members who might happen to be on Facebook. We thus expect that, to be successful, projects funded by occasional investors should be characterized by  a considerable number of Facebook friends. To see whether this is the case, we plot the probability that an investor supports a project as a function of the number of the project founder's Facebook friends (Figure~\ref{fig:group_numbacked_all_prob_facebookfriend}), and indeed find that projects whose founders have many Facebook friends tend to attract occasional investors, while founders with moderate numbers of Facebook friends attract frequent investors, partly confirming our expectation.

\begin{figure}[h!] \begin{center}
{\includegraphics[width=.35\textwidth]{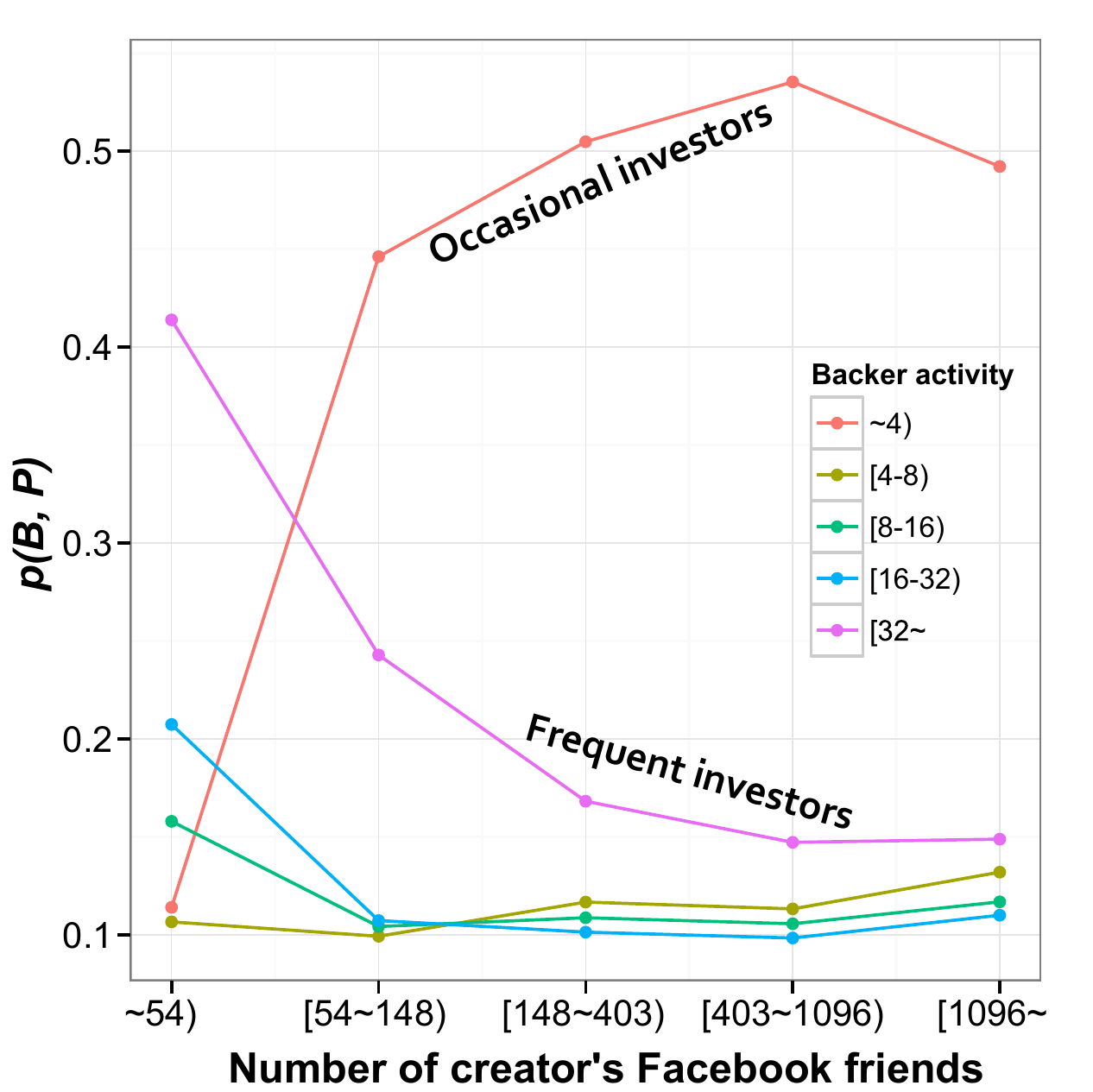} \label{fig:group_numbacked_all_prob_facebookfriend}}
\end{center} 
\caption{Probability that investor $B$ funds project $P$ as a function of  the number of the project founder's Facebook friends.} 
\label{fig:group_numbacked_all_prob_facebookfriend} \end{figure}

\section{Recommending Investors}
\label{sect:prediction} 

\noindent Based on the previous results, we are now able to recommend potential investors for a specific project. We do so by using logistic regression (LR)  and Support Vector Machine (SVM).  We use three different SVM kernels: linear, polynomial, and RBF (Radial Basis Function). The latter is more flexible and general than the first two as it copes with situations in which the relationships between features are non-linear.

To recommend potential investors who are on Twitter, we need to link Kickstarter users to their Twitter accounts first. We do so by matching the names of  Kickstarter users interested in a project with Twitter users mentioning the project. In so doing, we end up with 7,429 investors who are on Twitter, and  with 891 projects they had funded. To preliminarily test the accuracy of such matching, we randomly select 200 matching and manually inspect them: the resulting accuracy is  92\%.

\mbox{ }
\subsection{Experimental Setup}
\label{sect:experimental_setup}

\noindent We initially formulate the task of predicting who funds what as a binary classification problem. For each project-investor pair, we predict whether the investor supports the project (prediction is 1) or not (prediction is 0). That translates into having, for each project,  an unordered  list of Twitter users who are likely to fund it. 

We run those predictions on input of features that are both  \textit{static}  and \textit{dynamic}. Static features are permanently set at the start of the campaign and include a project's pledging goal, reward level, and category. They also include an investor's past supported project categories and his/her interests expressed on Twitter. Dynamic features, instead, change as  the campaign unfolds and include pledge growth rate, number of project updates, geographic dispersion of investors, and the number of comments exchanged between the founder and the community.

Since our data only includes positive cases -- that is, the set of pledges that actually happened -- we need to augment our dataset with negative cases (by under-sampling them): we do so by adding an equal number of negative cases -- that is, with a set of random  project-investor pairs. By construction, the resulting sample is balanced (the response variable is split 50-50), and the accuracy of a random prediction model would thus be 50\%. 

To evaluate the performance of the logistic regression and SVM without running into the problem of over-fitting, we perform 5-fold cross validation. That is, we randomly split the projects into two subsets: the first contains 80\% of the projects and is used for training; the second set contains 20\% of the projects and is used for testing. We repeat this split for 5 rounds and average the performance results across those rounds. 

As evaluation metrics,  we resort to those that are widely-used in classification problems: Accuracy (ACC), Precision, Recall, F-score, and Area Under the receiver-operator characteristic Curve (AUC).

\subsection{Experimental Results}

\noindent Before training any of the models, we compute the (Pearson) correlation coefficient between each pair of project features (Table~\ref{table:correlation}). We find that few features are correlated with each other (i.e., there are high positive correlations (where $r > 0.50$) between the pledging goal, the number of updates and the number of comments). Since it is not useful to simultaneously use all the features in the classification task, the input for the LR will include only the features that are not strongly correlated with each other (i.e., we only include the number of comments among those three features).\\

\begin{table}[t!]
\center\small\frenchspacing
\footnotesize
		\begin{tabularx}{0.9\linewidth}{r l l l l l l } 
	    & \rotatebox{90}{\mbox{\textbf{\#Updates}}} &  \rotatebox{90}{\mbox{\textbf{\#Comments}}} & \rotatebox{90}{\mbox{\textbf{Reward level}}} & \rotatebox{90}{\mbox{\textbf{Goal}}} & \rotatebox{90}{\mbox{\textbf{Growth rate}}} & \rotatebox{90}{\mbox{\textbf{Geographic dispersion}}} \\ \hline
	   \textbf{\#Comments} & \textbf{0.67} $\ast$ &  &  &  &  & \\
	   \textbf{Reward level} &0.12 & 0.03 &  &  & &  \\
	   \textbf{Goal} & \textbf{0.60} $\ast$ & \textbf{0.85} $\ast$  & 0.19 &  &  & \\
	   \textbf{Growth rate} & 0.34  & 0.12 & 0.33 & 0.11 & & \\
	   \textbf{Geo-D} & 0.12 & 0.21 & 0.16 & 0.23 & 0.13 & \\
	   \hline \\
	   \textbf{Activity level} & 0.26 & 0.19 & 0.05 & 0.21 & 0.32 & 0.17 \\
	\end{tabularx}
	\caption{Pearson correlation coefficients between each pair of features. Coefficients greater than $\pm 0.5$ with statistical significant level $<$ 0.05 are marked with a $\ast$.}
\label{table:correlation}
\end{table}

\begin{table}[t!]
\centering
\begin{center}
%\small 
%\frenchspacing
\hspace*{-5mm}
\begin{tabular}{c|c|ccccc}
\hline
\textbf{Model}  & \textbf{Features}  &  \textbf{ACC}   & \textbf{P} & \textbf{R}& \textbf{$F_1$}& \textbf{AUC}\\
\hline\hline
LR & Static  & 0.57 &  0.57 & 0.55 &  0.56 &  0.57\\
& Dynamic  & 0.57 &  0.58 & 0.55 &  0.56 &  0.57\\
\hline
SVM-linear & Static & 0.58 &  0.60 & 0.51 & 0.55 &  0.58\\
 & Dynamic & 0.58 & 0.60 & 0.50 & 0.55 &  0.58\\
\hline
SVM-poly & Static &  0.80 &  0.81 & 0.75 & 0.79 & 0.80\\
& Dynamic & 0.68  & 0.76 & 0.54 & 0.63 & 0.68\\
\hline
\textbf{SVM-RBF}  & Static & \textbf{0.82} &  0.79 & 0.83 & 0.82 & {0.81} \\
& Dynamic & \textbf{0.73} &  0.75 & 0.68 & 0.71 & {0.73} \\
\hline
\end{tabular}
\end{center}
\vspace*{-3mm}
\caption{Prediction results with the \textbf{balanced} test dataset (50/50 split).} 
\vspace*{-2mm}
\label{tab:svm-classification-model}
\end{table}

\begin{figure}[t!] \begin{center}
{\includegraphics[width=.4\textwidth]{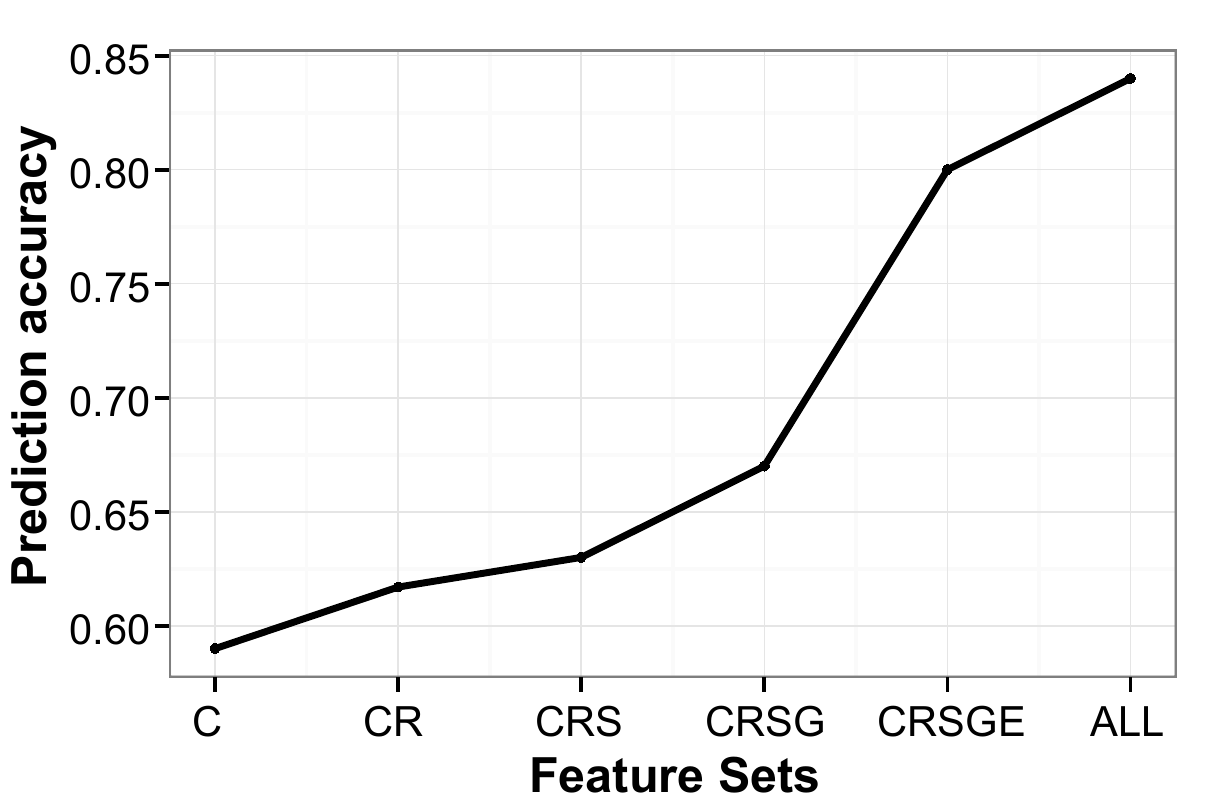}}
\end{center} 
\vspace*{-2mm}\caption{Model trained with different subsets of features} 
\label{fig:feature_selection} \end{figure}

\noindent \textbf{Prediction with balanced dataset.}  Table~\ref{tab:svm-classification-model} shows the results for our prediction models for balanced dataset on input of both static and dynamic features. We find that SVM with polynomial and RBF kernels work best, suggesting that our data points are not linearly separated in the hyperplane. Interestingly, on input of static features, the best classifier achieves 82\% accuracy (ACC); on input of dynamic features, the accuracy slightly degrades (73\%); while, as one expects, on input of both types of feature, the accuracy slightly increases to 84\%.

\begin{figure*}[t!] \begin{center}
\includegraphics[width=.9\textwidth]{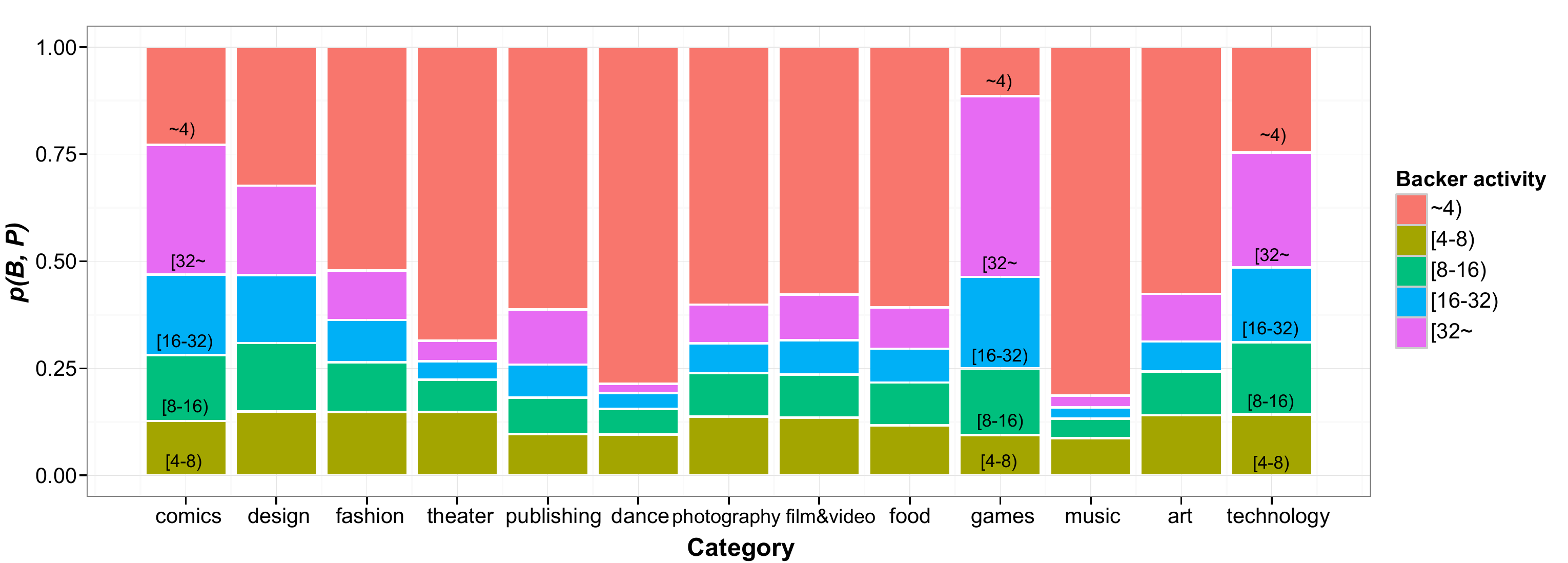}
\end{center} \vspace*{-5mm} \caption{Probability that investor \textit{B} (with a given level of activity) funds project {P} across different project categories.}\vspace*{-3mm}\label{fig:group_numbacked_all_prob_category_bar} \end{figure*}

These predictions are done upon the complete set of uncorrelated features. However, to know which feature individually matters and which does not, we re-run our classifications on input of different combinations of the following features: number of comments (C), reward levels (R), geographic span (S), growth rate (G), category matching (E) and topic similarity (TS). Again, we exclude two features: pledging goal and number of updates as they are both strongly correlated with the number of comments. Figure~\ref{fig:feature_selection}  shows the corresponding prediction accuracies: all features individually help to predict pledging behavior, but adding category matching and topical similarity results in considerable  performance improvements. We can then confirm that by visual inspection of Figure~\ref{fig:group_numbacked_all_prob_category_bar}. This plots the probability that an investor with a given level of activity funds a project in a given category. We see that occasional investors (red bar segments) support music projects, while active investors (purple bar segments) support projects with high pledging goals in, say, the gaming industry. \\

\noindent \textbf{Prediction with imbalanced dataset.} Our evaluation has so far assumed a 50-50 split between positive and negative cases. As this might not always be the case, we create an alternative test set with a 20/80 split (positive/negative): we train our models still on the balanced set but test them on the newly created unbalanced test set. We find that the results are similar to those obtained before (Table~\ref{tab:svm-classification-model-imbalanced}), yet with a minor degrade in precision. However, the accuracy (ACC) still remains as high as 82\%.

\subsection{Ranking investors}

\noindent To go beyond binary classifications, we now set out to \emph{rank} investors. So the problem is now, given a project, to return a ranked list of investors.

As evaluation metrics, we resort to two measures widely-used in ranking problems: MeanRR (Mean Reciprocal Rank) and MaxRR (Maximum Reciprocal Rank). We denote $rank_{i,p}$ the percentile-ranking of investor $i$ within the ordered list of investors predicted for project $P$: $rank_{i,P}$ = 0\% if  investor $i$  is predicted to be the most desirable for project $P$. Starting from this definition of rank, we can then formulate the total average percentile-ranking as:

\begin{equation}
{\overline{rank}} = {\frac{\sum\limits_{i,P}{funded_{i,P}} \dot {rank_{i,P}}}    {\sum\limits_{i,P}{rank_{i,P}}}}
\end{equation}

\noindent where $funded_{i,P}$ is a flag that reflects whether investor $i$ has supported project $P$: it is 0, if $i$ did not support it; otherwise, it is 1. The lower a list's $\overline{rank}$, the better the list's quality. For random predictions, the expected value for  $\overline{rank}$ is 0.5. Therefore, a $\overline{rank}$ < 0.5 is associated with an algorithm better than random. Given a ranked list, MeanRR returns the average rank score for the ``correct'' investors (i.e., investors who actually supported the project), while MaxRR returns the  score of the highest ranked correct investor (i.e., highest ranked among the investors who actually supported the project). The lower these two metrics, the better the ranking.

\begin{table}[t!]
\centering
\begin{center}
\begin{tabular}{c|c|cc}
\hline
\textbf{Model}   & \textbf{Features}  &  \textbf{ACC}   & \textbf{AUC}\\
\hline\hline
LR & Static   & 0.56 &   0.57\\
& Dynamic & 0.57 &   0.57\\
\hline
SVM-linear & Static & 0.60 &  0.58\\
& Dynamic & 0.61 &   0.59\\
\hline
SVM-poly & Static & 0.81 & 0.80\\
& Dynamic & 0.77 &   0.70\\
\hline
\textbf{SVM-RBF} & Static & \textbf{0.82} &   {0.81} \\
& Dynamic & \textbf{0.74} & {0.73} \\
\hline
\end{tabular}
\end{center}
\vspace*{-2mm}
\caption{Prediction results with the \textbf{imbalanced} test set (20/80 split).} 
\label{tab:svm-classification-model-imbalanced}
\vspace*{-2mm}
\end{table}

\begin{table}[t!]
\centering
\begin{center}
%\small 
%\frenchspacing
\hspace*{-5mm}
\begin{tabular}{c|c|cc}
\hline
 \textbf{Model} & \textbf{Features}  & \textbf{MeanRR} & \textbf{MaxRR} \\
\hline\hline
 Random & - & 0.50 & 0.87 \\
 \hline
 SVM-RBF & Static & 0.34 & 0.39 \\
  & Dynamic & 0.37 & 0.40 \\
  & \textbf{All} & \textbf{0.32} & \textbf{0.38} \\
\hline
\end{tabular}
\end{center}
\vspace*{-4mm}
\caption{Ranking Results.} 
\vspace*{-2mm}
\label{tab:recommending-model}
\end{table}

To ranking investors, we opt for the model that previously showed the best performance: SVM  with RBF kernel. This model returns the probability of the outcome to be ``1'' (i.e., the probability that an investor $B$ will support a project $P$). Upon our test set, for each project, we sort all users (a union set of training and test sets, which is the total number of 7,429 Twitter users) by this probability and recommend those who score the highest. In this way, it is similar to content-based recommendation. Table~\ref{tab:recommending-model} compares SVM's ranking accuracy with random model's. We can see that based only on static features, we can achieve percentile ranking of 0.34, which is lower than random model.

\section{Discussion}
\label{sect:discussion}

\noindent We now discuss the theoretical and practical implications of our work. 

\vspace*{-2mm}
\subsection{Theoretical Implications}

\noindent There is  a debate about the motives of crowdfunding investors. Initially, they were seen as donors~\cite{lehner@2013,ordanini@2011}: ``Some crowdfunding efforts, such as art or humanitarian projects, view their funders as patrons or philanthropists, who expect nothing in return.''~\cite{belleflamme@core2011}. In the same vein, Gerber \textit{et al.} listed the following motivations for investors: seek (non-financial) rewards, support creators and causes, engage and contribute to a trusting and creative community~\cite{gerber@cscw2012}. More recently however, crowdfunding sites have been increasingly attracting a variety of founders: from small entrepreneurs who traditionally relied on the 3Fs (friends, family, and fools~\cite{belleflamme@2010}) to big companies that now use those sites as marketing tools. In line with  these changes over the years, we have found that investors also tend to be of different types: pledging behavior of frequent  investors is very different from that of occasional ones. The former act as proper investors, while the latter act as donors. They generally support  different projects: art projects (e.g., music, dance) are largely funded by occasional investors, while projects on technology, games, and comics are funded by frequent ones. This suggests that pledging campaigns need to identify the right target investors to be successful. Artistic projects should rely on the traditional 3Fs (friends, family, and fools), perhaps employing social media sites to efficiently reach them. By contrast, technology projects should broaden their search and look for active and frequent investors.

\vspace*{-2mm}
\subsection{Practical Implications}
\noindent The good news is that we have shown that it is possible to identify and recommend frequent investors.  However, it might  not be sustainable to simply recommend investors only out of Kickstarter users: such an investor pool would limited and we could consequently end up recommending the same investors over and over again. To see whether we could expand the investor pool, it might be beneficial to study whether we could match unknown investors (who are not on Kickstarter but only on Twitter) with projects to be funded. To this end, we combine both static and dynamic \emph{project} features (Section~\ref{sect:experimental_setup}) with  \emph{Twitter}-derived features to test whether we can predict potential investors for each project. The Twitter-derived features are widely-used to measure activity, status~\cite{leskovec@www2010}, and influence~\cite{cha_icwsm10}, and they are three: 1) the logarithm of the total number of tweets (activity); 2)  the logarithm of the total number of followers divided by the number of followees (status); and 3) the sum of the average number of retweets, favorites, and mentions of the account's tweets (influence).
 
Using cross validation on our data in a way similar to what we have already done in Section~\ref{sect:experimental_setup}, we train  the SVM-RBF model, which  previously showed the best performance, solely on project and Twitter-derived  features. We learn that this model achieves ~68\% of accuracy (ACC in Table~\ref{tab:svm-classification-model-twitter}) and an average percentile ranking of 0.4 (Table~\ref{tab:recommending-model}), making it partly possible to recommend investors in cold-start situations and, as such, considerably expanding the investor pool.

\begin{table}[t!]
\centering
\begin{center}
%\small 
%\frenchspacing
%\hspace*{-5mm}
\begin{tabular}{c|c|ccccc}
\hline
\textbf{Model} & \textbf{Features}  & \textbf{ACC}  & \textbf{P} & \textbf{R}& \textbf{$F_1$}& \textbf{AUC}\\
\hline\hline
{SVM-RBF} & Static  & \textbf{0.68} & 0.71 & 0.61 & 0.66 & {0.68}\\
& Dynamic & \textbf{0.67} & 0.72 & 0.58 & 0.64 & {0.67} \\
\hline
\end{tabular}
\end{center}
\vspace*{-3mm}
\caption{Prediction results for Twitter+Project features.} 
\label{tab:svm-classification-model-twitter}
\end{table}

\begin{table}[t!]
\centering
\begin{center}
%\small 
%\frenchspacing
%\hspace*{-5mm}
\begin{tabular}{c|c|cc}
\hline
 \textbf{model} & \textbf{Features}  & \textbf{MeanRR} & \textbf{MaxRR} \\
\hline\hline
 Random & - & 0.50 & 0.87 \\
 \hline
 SVM-RBF & Static & 0.44 & 0.47 \\
  &Dynamic & 0.44 & 0.46 \\
  &\textbf{All} & \textbf{0.40} & \textbf{0.41} \\
\hline
\end{tabular}
\end{center}
\vspace*{-3mm}
\caption{Ranking results for Twitter+Project features.} 
\label{tab:recommending-model}
\end{table}

 \vspace*{-2mm}
\section{Conclusion}
\label{sec:conclusion}

\noindent Everyday there are on average 39  new projects in Kickstarter: not only artists or entrepreneurs are profiting from this new way of raising funds, but also city councils and political organizations have joined the fray. This is the first study to characterize the pledging behavior of micro-funders. We have established that investors behave quite differently depending on whether they are very active in the community or not. Frequent investors are attracted by  ambitious projects, yet they carefully diversify their investment portfolios. By contrast, occasional investors act as donors, mainly in art-related projects. 

We have also shown that it is possible to match new projects with willing investors, and that is extremely important, not least because the most common reason for failure in Kickstarter is the inability of founders to reach out to the right investors. 

We are currently working on a website that, on input of a Kickstarter project's url, will recommend a list of potential investors' Twitter accounts. We will then test the extent to which Kickstarter founders find this application useful. As for additional analysis, we are planning to look at exogenous factors, as this study has focused mainly on endogenous ones.

\vspace*{-2mm}
\section{Acknowledgment} 

\noindent Jisun An is supported in part by the Google European Doctoral Fellowship in Social Computing. We thank Nicola Barbieri, Martin Saveski, Alan Said, and Haewoon Kawk for their valuable comments on earlier versions of the  draft.  This work was done during an internship at Yahoo Labs in Barcelona. 

%EU SocialSensor FP7 project (contract no. 287975).

\balance

\vspace*{2mm}
\bibliographystyle{abbrv}

% ACM needs 'a single self-contained file'!
%

\balancecolumns
% That's all folks!

%\input{appendix.tex}
\end{document}